# Quasiparticle Energies and Excitonic Effects of Chromium Trichloride: from Two Dimensions to Bulk

Linghan Zhu[1], and Li Yang[1, 2]

[1]Department of Physics, Washington University in St. Louis, St. Louis, MO 63130, USA.

[2]Institute of Materials Science and Engineering, Washington University in St. Louis, St. Louis, MO 63130, USA.

**Abstract**

Layered van der Waals (vdW) magnetic materials have attracted significant research interest to date. In this work, we employ the first-principles many-body perturbation theory to calculate excited-state properties of a prototype vdW magnet, chromium trichloride ($CrCl_3$), covering monolayer, bilayer, and bulk structures. Unlike usual non-magnetic vdW semiconductors, in which many-electron interactions and excited states are sensitive to dimensionality, many-electron interactions are always enhanced and dominate quasiparticle energies and optical responses of both two-dimensional and bulk $CrCl_3$. The electron-hole (*e-h*) binding energy can reach 3 eV in monolayer and remains as high as 2 eV in bulk. Because of the cancellation effect between self-energy corrections and *e-h* binding energies, the lowest-energy exciton ("optical gap") is almost not affected by the change of dimensionality. Besides, for the excitons with similar *e-h* binding energies, their dipole oscillator strength can differ by a few orders of magnitude. Our analysis shows that such a big difference is from a unique interference effect between complex exciton wavefunctions and interband transitions. Finally, we find that the interlayer stacking sequence and magnetic coupling barely change quasiparticle band gaps and optical absorption spectra of $CrCl_3$. Our calculated low-energy exciton peak positions agree with available measurements. These findings give insight into the understanding of many-electron interactions and the interplay between magnetic orders and optical excitations in vdW magnetic materials.

## I. Introduction

Many-electron interactions are known to be enhanced in low-dimensional structures, and they dominate corresponding excited-state properties. Because of the reduced screening, electronic self-energy and excitonic effects are dramatically enhanced. For instance, the electron-hole (*e-h*) binding energy of monolayer transition metal dichalcogenides (TMDs) [1–7] and black phosphorus [8–11] can be around a few hundred meV, which is about one to two orders of magnitude larger than those of typical bulk semiconductors [1,7,9–13]. This reduced screening effect stems from the surrounded vacuum and, thus, is sensitive to dimensionality. For example,

*e-h* binding energy of bulk Tellurium is less than 10 meV while that of its monolayer structure is increased to be around 700 meV [14]. In addition to screening, the electronic band dispersion (effective mass) also impacts many-electron interactions. Particularly, flat bands contribute to a large joint density of states (JDOS), enhancing the chance of forming *e-h* pairs [15]. However, there have been very limited studies to clarify the roles of dimensionality and band-curvature effects on many-electron interactions and excitonic effects to date.

Recently achieved two-dimensional (2D) magnetic materials [16–19] may provide a unique opportunity to answer this fundamental question because their electronic band edges are usually dominated by localized 3*d* orbitals and are flat for both 2D and bulk structures. With the help of magnetic anisotropy [20], these 2D structures hold a long-range magnetic order by gapping low-energy modes of magnons [21]. Because of enhanced light-matter interactions, layered van der Waals (vdW) magnetic structures exhibit significant magneto-optical effects, such as the magneto-optical Faraday and Kerr effects and magnetic circular dichroism (MCD), which have been applied to identify magnetic orders [17,18,22] and give rise to optomagnetic applications based on vdW structures [23,24]. More recently, enhanced excitonic effects on magneto-optical responses have been calculated for monolayer magnetic structures [25,26]. These theoretical works showed that *e-h* binding energy can be 1.7 eV in monolayer $CrI_3$ [25], which is substantially larger than those of other non-magnetic semiconductors and satisfactorily explained available measurements.

Chromium trichloride ($CrCl_3$) is a good candidate to explore the relationship between many-electron interactions and dimensionality in correlated vdW magnetic structures. Unlike the widely studied $CrI_3$, whose bulk exhibits a ferromagnetic (FM) order [18,27], bulk $CrCl_3$ exhibits an interlayer antiferromagnetic (AFM) order with an intralayer FM order [19], the so-called A-type AFM order. Moreover, because of the significantly smaller spin-orbit coupling (SOC), the magnetic anisotropy energy (MAE) is small in $CrCl_3$ [19,21], and its 2D structure may exhibit an in-plane ground state magnetism [19,28,29] and rich topological spin textures, such as meron-like pairs at finite temperatures [30]. This weak SOC also substantially simplifies the optical spin selection rules, making $CrCl_3$ ideal to analyze many-electron interactions and corresponding optical activities.

In this work, we have employed first-principles many-body perturbation theory (MBPT) to study many-electron interactions and excited-state properties of $CrCl_3$. Significant self-energy corrections and excitonic effects are discovered for monolayer, bilayer and bulk $CrCl_3$. Our calculated exciton energies are in good agreement with available measurements for both monolayer and bulk structures. Interestingly, for excitons with similar *e-h* binding energies within the same structure, their dipole oscillator strength can be different for a few orders of magnitude. Our analysis shows that this big difference is from a unique interference effect between the complex exciton wavefunctions and interband transition matrices. Moreover, we show that many-electron interactions and excitonic effects are less sensitive to dimensionality in these vdW magnets. Thus, the flat bands and enhanced JDOS play an important role in deciding excited-state properties, giving hope to robust room-temperature excitons in bulk magnets. Finally, we find that these optical spectra and many-electron effects have little dependence on the interlayer crystallographic structure or magnetic orders. Therefore, magneto-optical effects [18,22] or second-harmonic generation (SHG) [31,32] may be applied to identify those complex symmetries of ultra-thin magnets.

The remainder of this article is organized as follows: In Sec. II, we introduce the atomic and crystallographic structure of $CrCl_3$, as well as the simulation setups. The results on the electronic and optical properties for monolayer, bilayer and bulk $CrCl_3$ are presented in Sec. III, IV and V respectively. In Sec. VI, we made a comprehensive comparison of the evolution of band gaps and optical absorbance with dimensionality. Finally, the conclusions are summarized in Sec. VII.

## II. Atomic Structures and Computational Details

The atomic structures of $CrCl_3$ are presented in Figure 1. Below 240 K [19], bulk $CrCl_3$ is a vdW layered material adopting the rhombohedral phase (space group $R\bar{3}$), which is formed by an interlayer shift along the $(\vec{a} - \vec{b})$ direction, as shown in Figures 1 (a1) and (a2). This is also known as the low-temperature (LT) phase. Above 240 K, it experiences a crystallographic phase transition into the monoclinic phase (space group *C2/m*), which is formed by an interlayer shift along the $-\vec{a}$ direction, as shown in Figures 1 (b1) and (b2). This is the high-temperature (HT) phase. Such a structural phase transition is similar to that observed in bulk $CrI_3$ [27]. Within each single layer of $CrCl_3$, the chromium atoms are arranged in a honeycomb structure, and each chromium atom is surrounded by six chloride atoms forming an octahedra. Below 17 K, an intralayer FM order is formed in bulk $CrCl_3$, followed by an interlayer AFM order below 14 K [19]. The magnetic moments are dominantly hosted on chromium atoms. The interlayer AFM/FM orders are schematically shown in the Figures 1 (a2) and (b2), respectively, with each layer taking the FM order within the layer.

The ground-state properties are obtained by density functional theory (DFT) within the general gradient approximation (GGA) using the Perdew-Burke-Ernzerhof (PBE) exchange-correlation functional [33] as implemented in the Quantum ESPRESSO package. The vdW interactions in bilayer and bulk $CrCl_3$ are included via the semiempirical Grimme-D3 scheme [34]. Semi-core 3*p* and 3*d* electrons of Chromium atoms are included in norm-conserving pseudopotentials [35], and the plane-wave energy cutoff is set to be 65 Ry. A vacuum distance of 18 Å between adjacent layers is used along the periodic direction in monolayer and bilayer calculations to avoid spurious interactions. SOC is relatively small in $CrCl_3$ due to the small atomic number of the ligand atom and, hence, is not considered in our calculations. [See supplementary information [36]]

The MBPT calculations are performed using the BerkeleyGW code [37] including the slab Coulomb truncation for monolayer and bilayer structures. Quasiparticle (QP) energy is calculated by using the single-shot $G_0W_0$ approximation within the general plasmon pole model [38]. The Bethe-Salpeter Equation (BSE) is employed to obtain excitonic effects and optical absorption spectra [39]. Because of the depolarization effect [40,41], only the incident light polarized parallel to the atomic plane (the in-plane direction) is considered for the monolayer and bilayer cases. 10 valence bands and 6 conductions bands are included in optical calculation to provide converged spectra below 3.5 eV. For monolayer and bilayer CrCl₃, a coarse *k*-grid of 9x9x1 is adopted to calculate the dielectric function and QP energies, and it is interpolated to a fine *k*-grid of 18x18x1 for the *e-h* interaction kernel and solving BSE. The coarse *k*-grid is set to be 9x9x2 and the fine *k*-gird is set to be 18x18x4 for bulk CrCl₃. The convergence test supplied in the supplementary information [36] shows that the error bar of this *k*-point sampling is within 0.1 eV for GW band gaps and exciton energies.

### III. Monolayer CrCl3

The DFT-calculated band structure of FM monolayer CrCl₃ is presented in Figure 2 (a). For FM monolayer CrCl₃, the valence band maximum (VBM) is located at the high-symmetry $M$ point and the conduction band minimum (CBM) is located around the middle point of the $K - \Gamma$ line, resulting in an indirect band gap of 1.84 eV and a direct band gap of 1.87 eV at $M$ point. According to the projected density of states (PDOS) in Figure 2 (a), the four lowest-energy conduction bands and the five highest-energy valence bands have a sizable amount of 3*p* orbital components of chloride atoms. Interestingly, those higher-energy (spin down) conduction bands between 2.5 and 3.5 eV are nearly from pure 3*d* orbitals of chromium atoms. Finally, both the valence and conduction band edges exhibit relatively flat dispersions and are composed of the same spin (up) states. This enhanced JDOS indicates strong many-electron interactions and potentially active interband transitions due to the spin-allowed selection rule [15].

We have calculated QP energies within the GW approximation. The DFT and GW band gaps are summarized in Table 1. The indirect QP band gap is 4.66 eV, and the direct QP band gap is 4.69 eV. The linear fit of GW-calculated quasiparticle energies to DFT eigenvalues is presented in Figure 2 (b). Due to the reduced dielectric screening of the suspended 2D structure and enhanced JDOS, QP energy corrections in monolayer CrCl₃ are significant, rendering a 2.82 eV enlargement of the GW band gap from the DFT result. This enhancement is larger than those in monolayer MoS₂ (~ 1.0 eV) [4,6] and black phosphorus (~ 1.2 eV) [10], indicating that correlated flat bands further enhance many-electron interactions in addition to the dimensionality factor.

We have further calculated the optical absorption spectrum of FM monolayer CrCl₃. To avoid the artificial effect from the choice of vacuum space in simulations, we plot the optical absorbance by $A(\omega) = \frac{\omega d}{c} \varepsilon_2(\omega)$, where $\varepsilon_2(\omega)$ is the calculated imaginary part of the dielectric function and $d$ represents the distance between adjacent CrCl₃ layers along the periodic direction of our calculation. Figure 3 (a) shows the optical absorbance of monolayer CrCl₃ with and without excitonic effects included. In the absence of *e-h* interactions, the absorbance (blue dashed line) edge starts at around 4.7 eV, corresponding to the QP direct band gap at the $M$ point. This significant absorption edge is consistent with the enhanced JDOS and allowed spin-selection rule as shown in Figure 2 (a).

After *e-h* interactions are included, the optical absorption spectrum is dramatically changed. The main optical absorption happens between 3 to 5 eV, which is a 2-eV red shift from the interband-transition result. Importantly, we observe two characteristic excitonic peaks in the optical spectrum at the low energy regime, which are marked as $X_1$ and $X_2$ at 1.48 eV and 2.25 eV in Figure 3 (a), respectively. It is worth mentioning that more excitonic states are around $X_1$ and $X_2$ while most of them are optically dark. In the inset of Figure 3 (a), we also mark the lowest-energy dark exciton, $X_0$, whose energy is about 20 meV below $X_1$ but its dipole oscillator strength is about four orders of magnitude smaller than that of $X_1$. These low-energy excitons result in significant exciton binding energies of 3.23 eV, 3.21 eV, and 2.44 eV for $X_0$, $X_1$, and $X_2$, respectively. Such exciton binding energies are enormous compared with those of other typical 2D semiconductors such as monolayer MoS$_2$ (~ 960 meV) [4] and black phosphorous (~ 800 meV) [10], and they are almost two times larger than that of the sibling magnetic material, monolayer CrI$_3$ (~ 1.7 eV) [25]. The *e-h* binding energy of the lowest bright $X_1$ exciton is even larger than the self-energy (GW) correction. As a result, the "optical gap" (1.48 eV) is lower than the DFT band gap (1.84 eV). Recent photoluminescence measurements of monolayer and multilayer CrCl$_3$ at 2 K found a single peak around 1.43 eV [28]. This agrees with our MBPT results, where the $X_1$ exciton is located at 1.48 eV.

We have tested the dependence of MBPT results on top of the DFT ground state with Hubbard potential, denoted by DFT+U/MBPT. We choose the Hubbard parameters [43] U=1.5 eV and J=0.5 eV as an example [26]. To avoid the double-counting problem in this DFT+U/MBPT scheme, we have subtracted the DFT+U-level Hubbard potential $V_{Hub}$ together with the DFT exchange-correlation potential $V_{xc}$ from the conventional self-energy operator $\Sigma(E)$ within the GW approximation, following Ref. [25,44,45] by:

$$[T + V_{ext} + V_H + V_{xc} + V_{Hub}]\Psi(\boldsymbol{r}) + \int d\boldsymbol{r}' \Delta\Sigma(\boldsymbol{r},\boldsymbol{r}';E^{qp})\Psi(\boldsymbol{r}') = E^{qp}\Psi(\boldsymbol{r}), \quad (1)$$

and

$$\Delta\Sigma(E) = \Sigma(E) - V_{xc} - V_{Hub}. \quad (2)$$

The DFT+U/MBPT absorption spectra of monolayer CrCl$_3$ are shown in Figure 3 (b). (See the supplementary information [36] for the DFT+U level band structure and PDOS.) After Hubbard potential is included, the DFT+U level band gap is around 150 meV larger than the DFT band gap, and the GW quasiparticle indirect gap is increased by about 180 meV to 4.84 eV, as seen from the onset of the absorbance without *e-h* interaction in Figure 3 (b). The optical absorption spectrum from DFT+U ground state also shows a significant blue shift. For example, the $X_1$ exciton energy is increased by around 400 meV to 1.87 eV. Given the better correspondence with available measurements [28] in the absence of Hubbard potential, we use the DFT/MBPT scheme without U in the following calculations of bilayer and bulk CrCl$_3$.

Moving to higher excitation energies (between 3 eV to 5 eV), there are exciton states with much stronger dipole oscillator strength than those lower-energy exciton states. For example, the peak marked by $X_3$ in Figure 3 (a) has an oscillator strength two orders larger than that of $X_1$. These bright excitons dominate the main optical absorption spectrum.

To better understand these strongly bound excitons, we have plotted the real-space exciton wavefunctions of $X_0$, $X_1$ and $X_3$ with the hole positioned on a chromium atom in Figures 4 (b)-(d). The choice of the location of the hole is decided by that band-edge valence states are dominated by the d orbitals of chromium atoms. Thus, setting the hole at chromium atoms will substantially reduce numerical noise of the plots. Because of the large *e-h* binding energy, all three excitons exhibit highly localized wavefunctions. Particularly, for $X_0$ and $X_1$, their wavefunctions are nearly confined within one unit cell. These highly localized real-space wavefunctions indicate a smearing of the *e-h* pair contributions from the whole Brillouin zone (BZ) in reciprocal space. This also agrees with the argument that those flat bands around band edges actively contribute to the formation of strongly bound excitons. For $X_3$, because of its smaller *e-h* binding energy, the wavefunction is slightly broader and roughly covers the size of three unit cells.

It is hard to tell any significant difference from the real-space wavefunctions of excitons $X_0$ and $X_1$, whose dipole oscillator strength differ, however, by four orders of magnitude. Following Ref. [46], we try to find the original contributions of dipole oscillator strength of these excitons. We rewrite the optical transition matrix element $\langle 0|\hat{v}|i\rangle$ from the ground (vacuum) state $|0\rangle$ to an exciton state $|i\rangle = \sum_{vck} A^i_{vck} |vc\rangle$ ($A^i_{vck}$ is the exciton amplitude solved from the BSE [39] and the interband transition happens between same spins because of the spin selection rule) to analyze the contribution of interband transition matrix elements $\langle vk|\hat{v}|ck\rangle$ at a certain energy $\omega$ to the optical transition matrix element

$$\langle 0|\hat{v}|i\rangle = \sum_{vck} A^i_{vck} \langle vk|\hat{v}|ck\rangle = \int S_i(\omega)\, d\omega, \tag{3}$$

where

$$S_i(\omega) = \sum_{vck} A^i_{vck} \langle vk|\hat{v}|ck\rangle \delta(\omega - (E_{ck} - E_{vk})), \tag{4}$$

and

$$I_i(\omega) = \int_0^\omega S_i(\omega')\, d\omega'. \tag{5}$$

The corresponding interference effect between the complex interband transition matrices ($\langle vk|\hat{v}|ck\rangle$) and exciton amplitude ($A^i_{vck}$) is essential for determining the overall dipole oscillator strength of excitons (note that the exciton dipole oscillator strength is proportional to the square of $I_i(\omega)$). Since monolayer CrCl$_3$ is FM and lacks time reversal symmetry, $S_i(\omega)$ and its integral $I_i(\omega)$ are complex functions. To address main characters, we only plot the real part of $S_i(\omega)$ and $I_i(\omega)$ in Figures 4 (e)-(g) for the exciton states $X_0$, $X_1$ and $X_3$, respectively. The imaginary part is similar. These plots essentially show how *e-h* interactions obtain dipole oscillator strength from interband transitions at different energies and reform them into corresponding excitons. Like previous studies on graphene [46], the energy distribution of $S_i(\omega)$ of all studied excitonic states is spread over a wide energy range, which is consistent with their large binding energies.

For the dark exciton $X_0$, there is a coherent cancellation of $S_i(\omega)$. As shown in Figure 4 (d), $S_i(\omega)$ fluctuates positively and negatively with similar amplitude at all energies. As a result, the integral of $S_i(\omega)$ is not able to build up over the energy range and produces a small overall $I_i(\omega)$. This is the reason for the tiny dipole oscillator strength of $X_0$. This interference effect between interband transition matrix elements and exciton wavefunctions is less prominent for the bight exciton state $X_1$, especially at the low energy side. As shown in Figure 4 (e), $I_i(\omega)$ grows dominantly from the quasiparticle band gap around 4.7 eV, and nearly saturates after 5 eV. This indicates that the dipole oscillator strength of the bright exciton $X_1$ is mainly contributed by those flat bands around band edges. Finally, for the bright exciton $X_3$ in Figure 4 (f), there is only minor interference effect. Particularly, the interband contributions have nearly the same positive sign over the whole energy range. As a result, the integral $I_i(\omega)$ builds consistently along the way to higher transition energies, resulting in a large oscillator strength. In a word, the dramatically different optical dipole oscillator strength of excitons with similar *e-h* binding energy is mainly from the interference effect between the complex interband transitions and exciton amplitude involved in forming excitons.

**IV. Bilayer CrCl$_3$**

Compared with monolayer CrCl$_3$, the interlayer magnetic order and stacking sequence in bilayer CrCl$_3$ bring more degrees of freedom. First, as shown in Figure 1, there are two crystallographic structures of bulk CrCl$_3$. Unlike bulk CrCl$_3$, recent experiments have shown that no crystallographic transition was observed in few layer CrCl$_3$, keeping a monoclinic HT phase structure at low temperatures [47]. Nevertheless, in order to identify the possible influence of crystallographic structure on the electronic and optical properties of bilayer CrCl$_3$, we consider both stacking sequences in our calculations. Second, the energy difference of interlayer FM and AFM orders is small, less than 5 meV/Cr as shown in previous calculations [19,42] and depends critically on the approach used [42]. Although interlayer AFM order is widely observed in available measurements [19,28,29,47], the FM order can be easily achieved by applying a small magnetic field as shown by recent experiments [19,28,29]. Therefore, we will consider both interlayer AFM and FM orders in the following calculations of bilayer CrCl$_3$.

The DFT-calculated band structures are summarized in Figure 5 for AFM/FM and LT/HT configurations of bilayer CrCl$_3$. For the interlayer AFM order shown in Figures 5 (a) and (c), the bands of the two layers are almost degenerate, with opposite spin components from each layer. For the interlayer FM order shown in Figures 5 (b) and (d), the two layers have same spin components, resulting in an overall double-degenerated spin up band edge states. Meanwhile, the band structures and band gaps of LT and HT phases are similar, except that the monoclinic HT structure exhibits a little larger splitting of the bands than the rhombohedral LT structure.

The GW-calculated QP energy vs the DFT results are summarized in Table 1. Because of the similar reasons of enhanced self-energy corrections as in monolayer CrCl$_3$, significant QP energy corrections are obtained in bilayer CrCl$_3$. For example, the QP band gap of bilayer rhombohedral LT phase AFM CrCl$_3$ is increased from a DFT value of 1.84 eV to 4.45 eV. As shown in Table 1, these self-energy corrections are not sensitive to the interlayer structure or magnetic order; the self-energy corrections are around 2.6 eV for all four configurations. On the other hand, this GW correction is around 200 meV smaller than that of monolayer. This is mainly from the increased screening in bilayer structures.

The optical absorption spectra of these four configurations are presented in Figure 6. Like the results of QP energies, the optical absorption spectra are similar for all four configurations. Therefore, many-electron effects and linear optical absorption spectra are not sensitive to the interlayer crystallographic structure and magnetic orders. Take the bilayer LT phase of AFM $CrCl_3$ as an example (Figure 6 (a)). Without *e-h* interactions, the absorption edge starts at around 4.5 eV, which is due to the QP band gap. After including *e-h* interactions, a significant red shift of the optical spectrum is observed. Like the result of monolayer, the main optical spectrum is still located between 3 to 5 eV although the QP band gap is reduced by around 200 meV compared with monolayer. Moreover, the two characteristic excitonic peaks, $X_1$ and $X_2$, are observed in all spectra. Their energies are similar for all four configurations as well. Interestingly, the energies of $X_1$ are slightly higher than that of monolayer. As seen from Table 1, the GW correction for AFM rhombohedral (LT) bilayer is 2.61 eV, which is about 200 meV smaller than that of monolayer. However, the *e-h* binding energy of bilayer is reduced by about 300 meV than that of monolayer. As a result, the absolute value of exciton energy is slightly increased finally. This is an opposite trend according to usual quantum confinement effect, in which thinner samples show a blue shift of the optical spectrum [10,11,14]. Such an unusual confinement effect is due to the enhanced excitonic effects in $CrCl_3$, where *e-h* binding energy is larger than the self-energy (GW) correction. Following similar scaling law as dielectric screening increases, the reduction of *e-h* binding energy in thicker samples is larger than the reduction of self-energy correction. As a competition result, the absolute energy of the $X_1$ exciton is slightly increased. In other words, this unusual quantum confinement effect is essentially from the flat bands with significant joint density of states (JDOS). This quantum confinement effect was reported in previous studies of other vdW magnetic structures [25].

**V. Bulk $CrCl_3$**

We have performed the GW-BSE calculations on bulk $CrCl_3$. Given the results from bilayer $CrCl_3$ that quasiparticle energy corrections and absorption spectra are not sensitive to the interlayer stacking and magnetic order, we only consider the experimentally observed rhombohedral bulk structure with the AFM interlayer coupling [19] (see supplementary [36] for the results on the monoclnic strutcure). The DFT-calculated band structure of bulk $CrCl_3$ is presented in Figure 7 (a). Interestingly, quantum confinement effects are nearly negligible within DFT results: the band structure and band gap of bulk $CrCl_3$ is nearly the same as those of monolayer and bilayer structures.

The GW-calculated QP energy vs the DFT results are summarized in Table 1. Significant self-energy corrections are observed in bulk $CrCl_3$. Because of stronger screening in three dimensions, the GW enlargement of the band gap is around 2.0 eV, and it is smaller than those in monolayer

(around 2.8 eV) and bilayer (around 2.6 eV). Nonetheless, this reduction of band gap from monolayer (~ 4.66 eV) to bulk (~ 3.87 eV) is significantly smaller than other typical semiconductors such as black phosphorous (~ 2 eV in monolayer and less than 0.3 eV in bulk) [10] and tellurium (~ 2.35 eV in monolayer and less than 0.41 eV in bulk) [14].

Further we have calculated the absorption spectrum $\varepsilon_2(\omega)$ of bulk $CrCl_3$. Because the depolarization effect is negligible in bulk structures, we consider both in-plane and out-of-plane polarizations of incident light, as shown in Figures 7 (b) and (c), respectively. As expected, before including *e-h* interactions, both the optical absorption spectra start from the QP band gap around 3.9 eV. Excitonic effects substantially shift the main optical absorption spectrum to between 3 eV and 4 eV. For in-plane polarized incident light (Figure 7 (b)), those two characteristic excitonic peaks ($X_1$ and $X_2$) are similar to the monolayer and bilayer cases and are located at 1.78 eV and 2.51 eV with *e-h* binding energies of 2.11 and 1.38 eV, respectively. These exciton energies are higher than those of bilayer (1.55 and 2.31 eV) and monolayer (1.48 and 2.25 eV), exhibiting an opposite trend of the usual quantum confinement effects according to the same reason as explained in the Sec. IV for bilayer $CrCl_3$. Moreover, the dipole oscillator strength of these two characteristic peaks is also enhanced. This can be from the stronger interlayer hybridization that enhances the overlap of electron and hole wavefunctions and corresponding transition matrices.

The dipole oscillator strength distribution is largely different for different incident-light polarizations, resulting in a highly anisotropic optical spectrum. For incident light polarized along the out-of-plane direction (Figure 7(c)), the characteristic peak $X_2$ becomes optically dark, and the dipole oscillator strength of $X_1$ is further enhanced. Besides, the main absorption between 3 eV and 4 eV becomes more isolated absorption peaks around 3.1 eV and 3.8 eV. The redistribution of exciton dipole oscillator strength under different incident light polarization may be employed in experiments to detect the crystal orientation.

There are extensive experiments on the absorption spectrum of bulk $CrCl_3$, as summarized in Table 1. In both Refs. [19,48], they reported absorption peaks around 1.7 eV and 2.3 eV for bulk $CrCl_3$. These are in good accordance with our calculated absorption peaks at 1.78 eV ($X_1$) and 2.51 eV ($X_2$) and their energy splitting (0.73 eV). It has to be pointed out that the measurements of Ref. [48] were performed at 80 K and 300 K, which are above the Néel temperature (14 K) of bulk $CrCl_3$. It could be a problem to compare our results under a perfect AFM order to measurements of the paramagnetic order. Unfortunately, we cannot find optical measurements of bulk $CrCl_3$ under its Néel temperature. On the other hand, there is a report of another A-type AFM material $CrPS_4$, in which the photoluminescence peak positions are not shifted when passing the Néel temperature [49], although the peak width and shape change as temperature increases. Therefore, we expect that those absorption peaks in Refs. [19,48] will not be substantially changed by the magnetic order.

**VI. Evolution of band gaps and excitons with dimensionality**

Finally, we have summarized the evolution of band gaps and characteristic excitons of CrCl$_3$ from monolayer, bilayer, to bulk. Given available measurements and the insensitivity of electronic and optical properties to the interlayer stacking and magnetic configurations, we use the results of interlayer LT structures and AFM coupling, and the fitting results are universal for all configurations. In Figure 8 (a), the evolution of the DFT and QP band gaps as well as the "optical gap" (the first bright exciton peak $X_1$) is presented. To quantitatively provide the band gap dependence on the layer number, we employ the widely used empirical power law formula [10,11,50]:

$$E^N = E_{bulk}^\infty + \frac{A}{N^\alpha}, \qquad (6)$$

where $N$ is the layer number and $E_{bulk}^\infty$ represents the bulk value. The fitted results are included in Table 2. Although the DFT band gap barely changes from monolayer to bulk, the significant QP energy corrections reflect the trend of the increased dielectric screening effect. The failure of DFT in predicting the band gap as well as the dielectric screening effect indicates it is important to go beyond DFT in calculating the electronic properties of vdW layered magnetic materials even for obtaining qualitative trends of quantum confinement. Interestingly, the GW-calculated band gap follows the $1/N^{0.5}$ power law. This decay is slower than the usual quantum confinement case with $1/N^2$ [50] based on free-electron gas, and indicates that many-electron correlations are significantly less sensitive to the quantum confinement in correlated CrCl$_3$.

In Figure 8 (b), we focus on those two characteristic peaks ($X_1$ and $X_2$). As noticed in above presentations, the brightness of these two excitons are sensitive to the layer number of structures. As shown in Figure 8 (b), the absorbance of the lower-energy exciton ($X_1$) is more sensitive to the thickness, and it is increased from 0.06% in monolayer to 0.25% in bulk. Thus, we expect these two characteristic excitons can be useful to estimate the thickness of samples.

## VII. Summary

In summary, we have systematically studied the electronic and optical properties from monolayer, bilayer to bulk CrCl$_3$ using first-principles MBPT approach. Unlike typical semiconductors, the increased dielectric screening in bulk CrCl$_3$ only renders a less than 20% decrease in the QP band gap relative to the monolayer case, and the energy of the lowest bright exciton is even slightly increased from monolayer to bulk. Besides, the absorption spectrum of bulk CrCl$_3$ resembles that of the monolayer, with significant *e-h* binding energy of lowest exciton state around 2 eV compared with 3 eV in monolayer CrCl$_3$. The physics origin of different dipole oscillator strengths between excitons is discussed based on the interference effect between exciton wavefunctions and interband transition matrices. Our calculated results are in good agreements with available measurements. Finally, we find that the absorption spectra of the vdW magnet CrCl$_3$ is not sensitive to the interlayer magnetic order or stacking structure. Magneto-optical probes such as Kerr effect and MCD may be needed in future experiments to probe the magnetic order in these magnetic materials.

**Acknowledgment**

The work is supported by the National Science Foundation (NSF) CAREER Grant No. DMR-1455346 and the Air Force Office of Scientific Research (AFOSR) Grant No. FA9550-17-1-0304. The authors would like to thank Meng Wu, Prof. Yan Lyu, Xiaobo Lu and Ruixiang Fei for fruitful discussions. The computational resources have been provided by the Stampede of TeraGrid at the Texas Advanced Computing Center (TACC) through XSEDE.

**Tables:**

**Table 1** Summary of DFT and GW band gaps (the values listed are for the direct band gap, and the values in the parenthesis are for the indirect band gap), $X_1$ and $X_2$ exciton energy and their experimental values for monolayer, bilayer and bulk CrCl$_3$. The unit is eV.

|  |  | DFT band gap | GW band gap | $X_1$ energy | $X_1$ energy expt. | $X_2$ energy | $X_2$ energy expt. |
|---|---|---|---|---|---|---|---|
| Monolayer |  | 1.87 (1.84) | 4.69 (4.66) | 1.48 | ~ 1.43 (PL) [28] | 2.25 | – |
| 2L | Rhombohedral AFM | 1.87 (1.84) | 4.48 (4.45) | 1.55 | – | 2.31 | – |
|  | Rhombohedral FM | 1.83 (1.80) | 4.43 (4.40) | 1.53 | – | 2.29 | – |
|  | Monoclinic AFM | 1.85 (1.84) | 4.44 (4.43) | 1.54 | – | 2.29 | – |
|  | Monoclinic FM | 1.81 (1.80) | 4.41 (4.40) | 1.53 | – | 2.30 | – |
| bulk | Rhombohedral AFM (in-plane polarization) | 1.87 (1.85) | 3.89 (3.87) | 1.78 | ~ 1.7 (Abs) [19,48] | 2.51 | ~ 2.3 (Abs) [19,48] |

**Table 2** Fitting parameters of DFT, QP band gaps, and "optical gap" (energy of the first bright exciton $X_1$) to the layer number according to the power law formula $E_{bulk}^{\infty} + A/N^{\alpha}$.

|  | DFT | QP | "Optical gap" |
|---|---|---|---|
| $\alpha$ | 0.03 | 0.51 | 0.48 |
| A | -0.01 | 0.79 | -0.30 |

**Figures:**

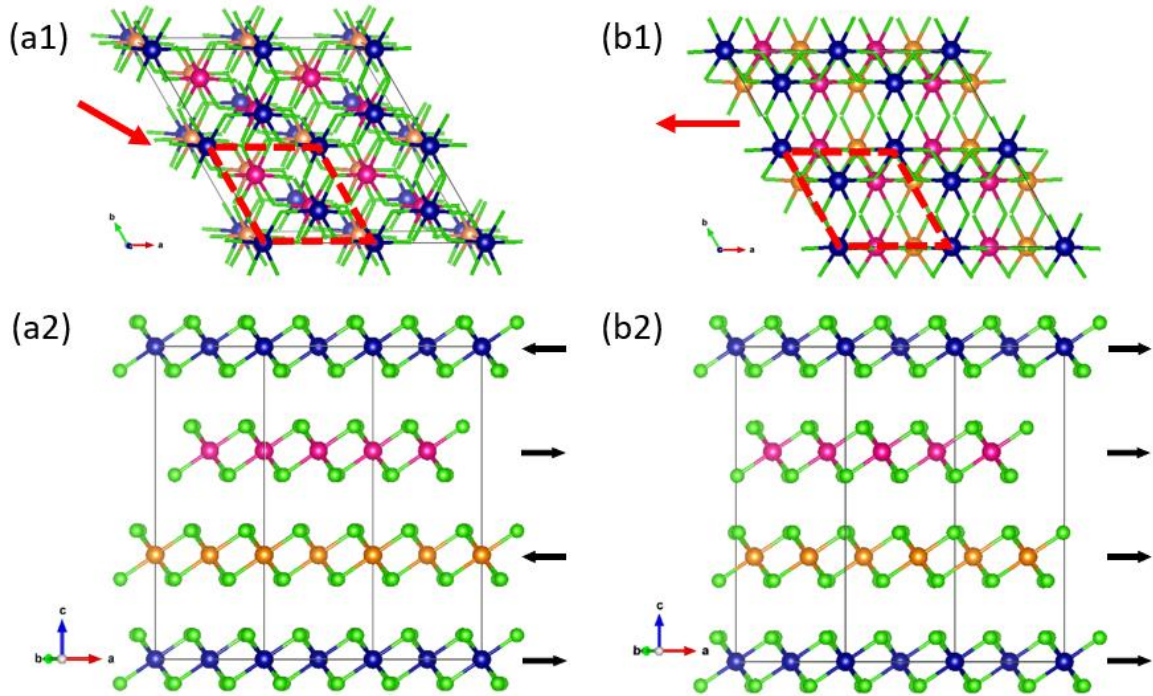

**Figure 1** (a1) and (a2) Top and side views of the rhombohedral (LT) structure. (b1) and (b2) top and side views of the monoclinic (HT) structure. Cr atoms in different layers are shown with different colors. The Cl atoms are not shown in top views ((a1) and (b1)) for clarity. The arrows in (a1) and (b1) show the relative interlayer shift direction. The structure in (a1) is slightly shifted along the interlayer shift direction for clarity. The unit cells are marked by red dashed lines. The interlayer AFM/FM orders are schematically shown in (a2) and (b2), respectively, by the black arrows representing atomic spin directions in each intralayer FM ordered layer. The polarizations are shown in the in-plane direction to reflect the experimentally observed in-plane magnetic polarizations, although only collinear polarizations are considered in the calculations.

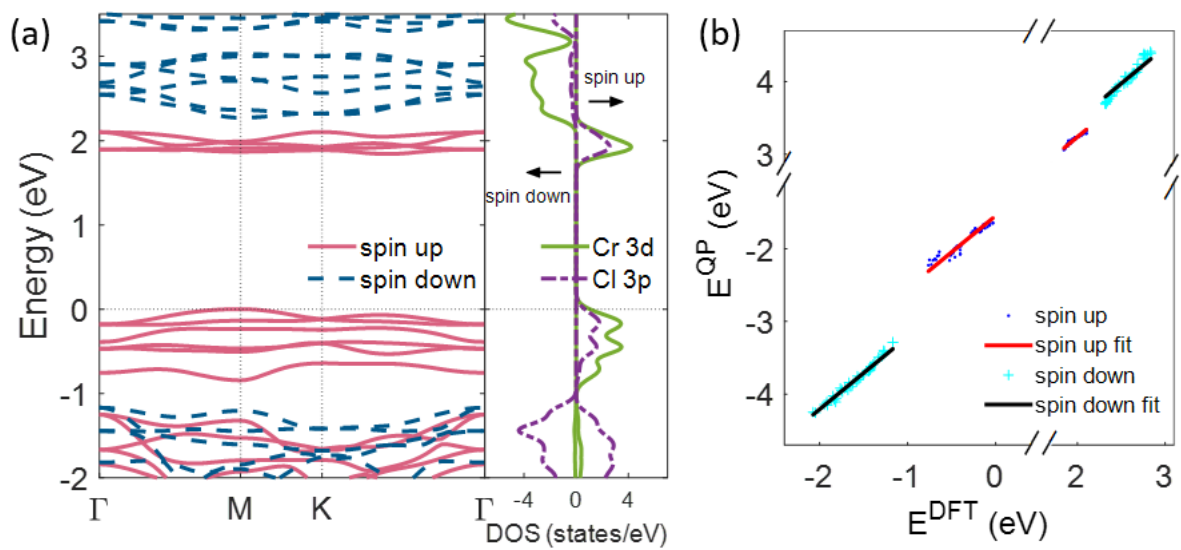

**Figure 2** (a) DFT-calculated band structure (left panel) and projected density of states (right panel) of monolayer CrCl$_3$ with an intralayer FM order. The energy of the valence band maximum is set to be zero. (b) Linear fit of QP energies to DFT eigenvalues for monolayer CrCl$_3$.

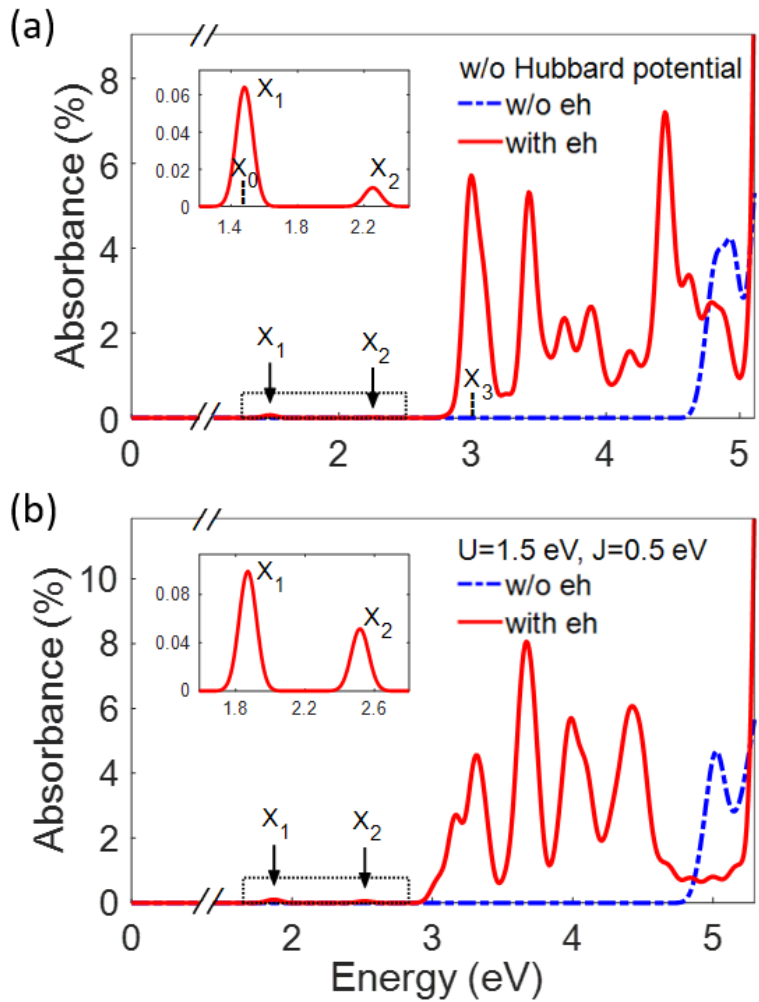

**Figure 3** Absorbance of monolayer CrCl$_3$ without *e-h* interaction (blue dashed line) and with *e-h* interaction (red solid line) calculated on top of (a) DFT ground state and (b) DFT+U ground state (with Hubbard parameters U=1.5 eV, J=0.5 eV). The two characteristic absorption peaks are marked as $X_1$ and $X_2$, respectively. A dark exciton state below $X_1$ is marked as $X_0$ in the inset of (a). An exciton state at higher energy is marked as $X_3$ in (a).

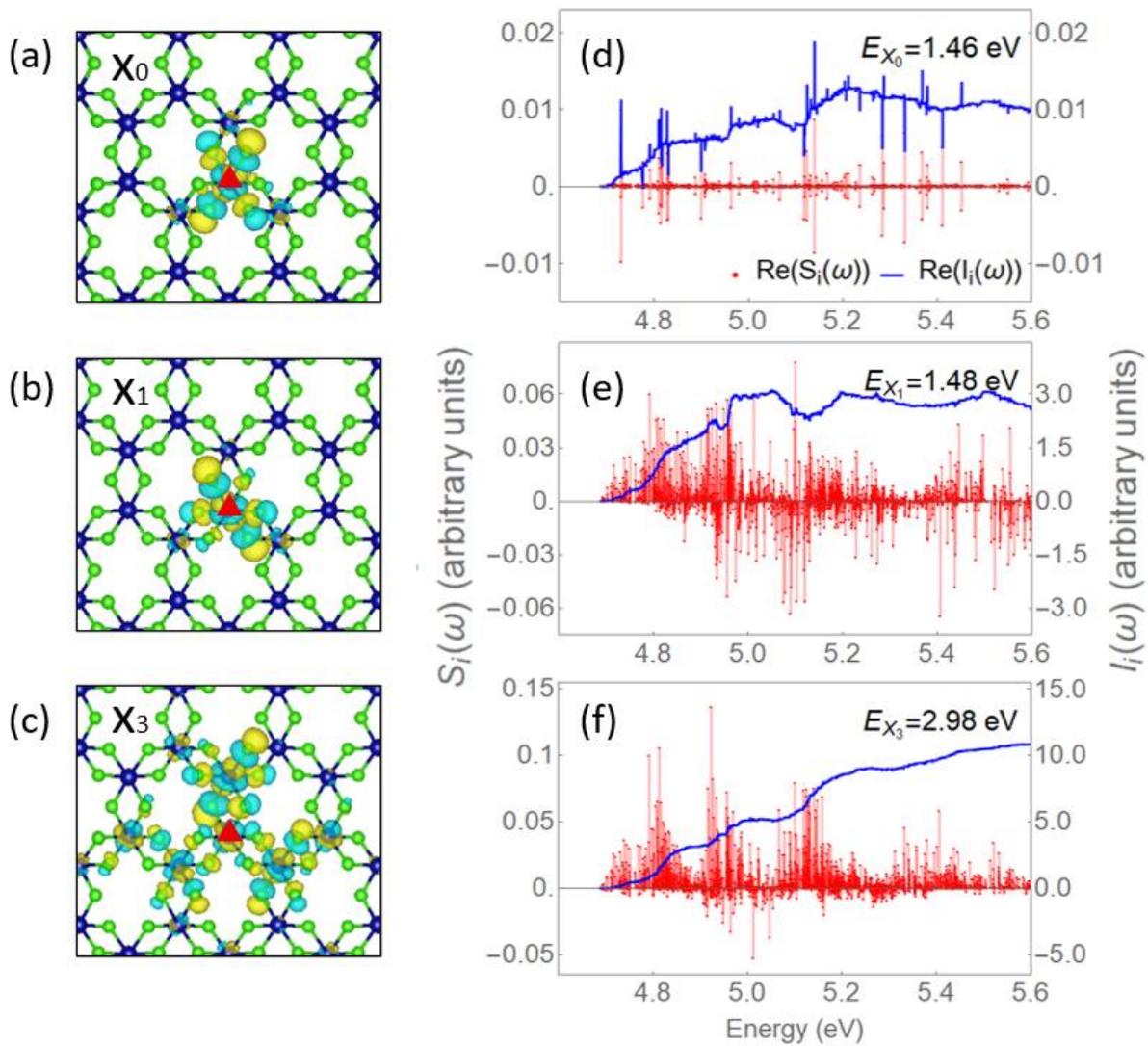

**Figure 4** (a-c) Real-space wavefunctions of exciton states $X_0$, $X_1$ and $X_3$ in monolayer CrCl$_3$, respectively. The hole positions are marked with red triangles, which are at Cr atoms. (d-f) $S_i(\omega)$ and its integral $I_i(\omega)$ for exciton states $X_0$, $X_1$ and $X_3$, respectively.

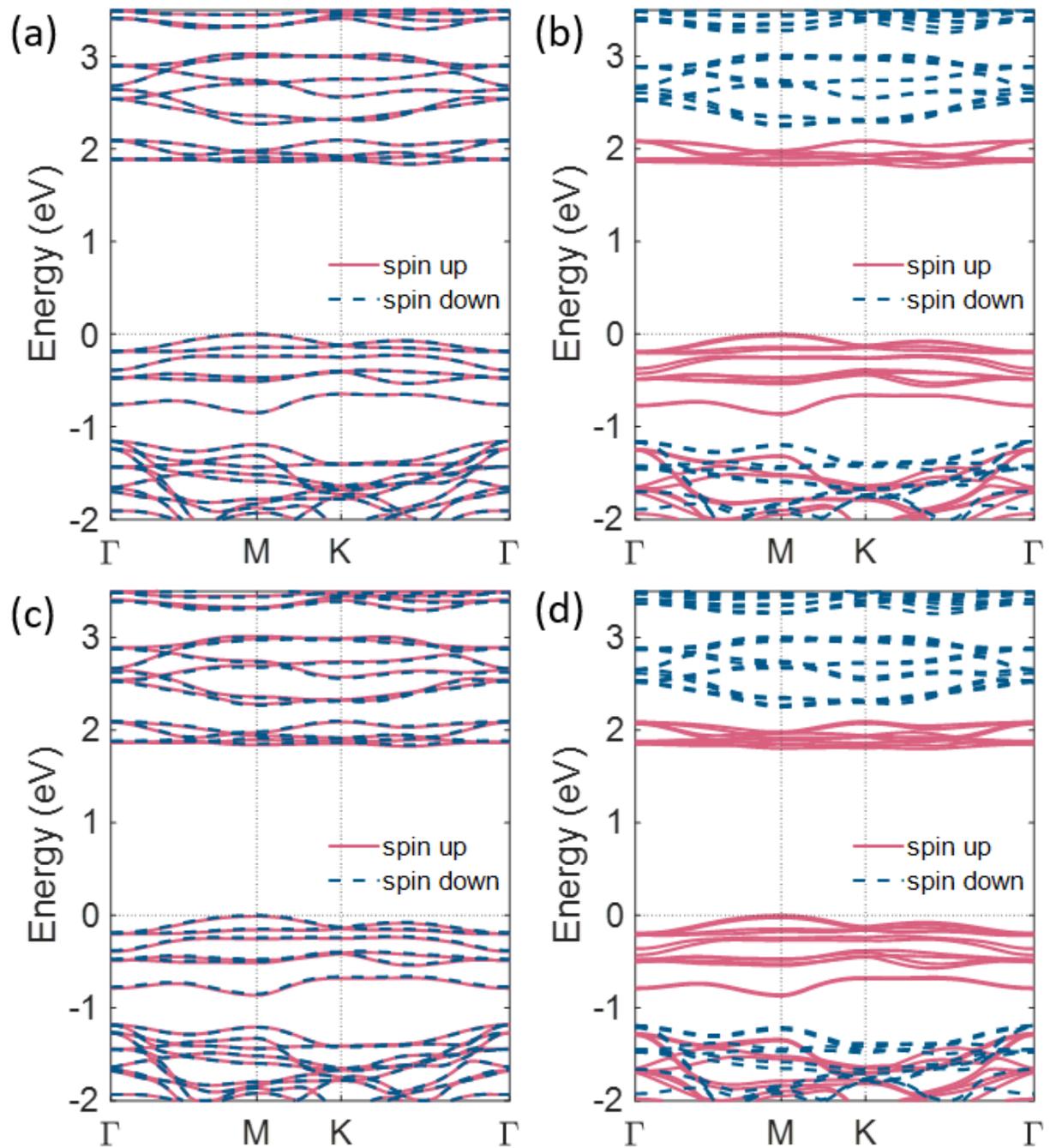

**Figure 5** DFT-calculated band structure of bilayer $CrCl_3$: (a) LT rhombohedral stacking with an AFM interlayer coupling; (b) LT rhombohedral stacking with a FM interlayer coupling; (c) HT monoclinic stacking with an AFM interlayer coupling; (d) HT monoclinic stacking with a FM interlayer coupling. The energy of the valence band maximum is set to be zero.

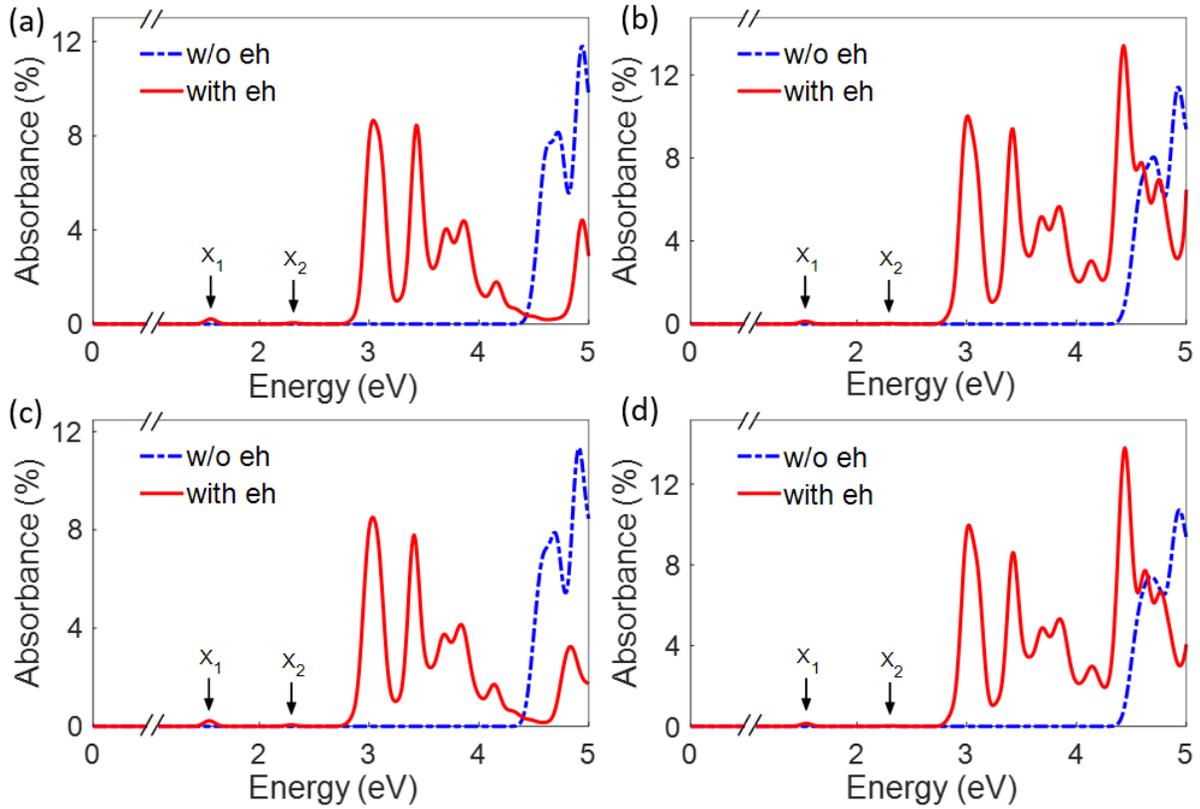

**Figure 6** Optical absorbance of bilayer CrCl$_3$ without and with *e-h* interactions: (a) those of the LT rhombohedral stacking with an AFM interlayer coupling; (b) those of the LT rhombohedral stacking with a FM interlayer coupling; (c) those of the HT monoclinic stacking with an AFM interlayer coupling; (d) those of the HT monoclinic stacking with a FM interlayer coupling. The two characteristic excitonic peaks are marked as $X_1$ and $X_2$, respectively.

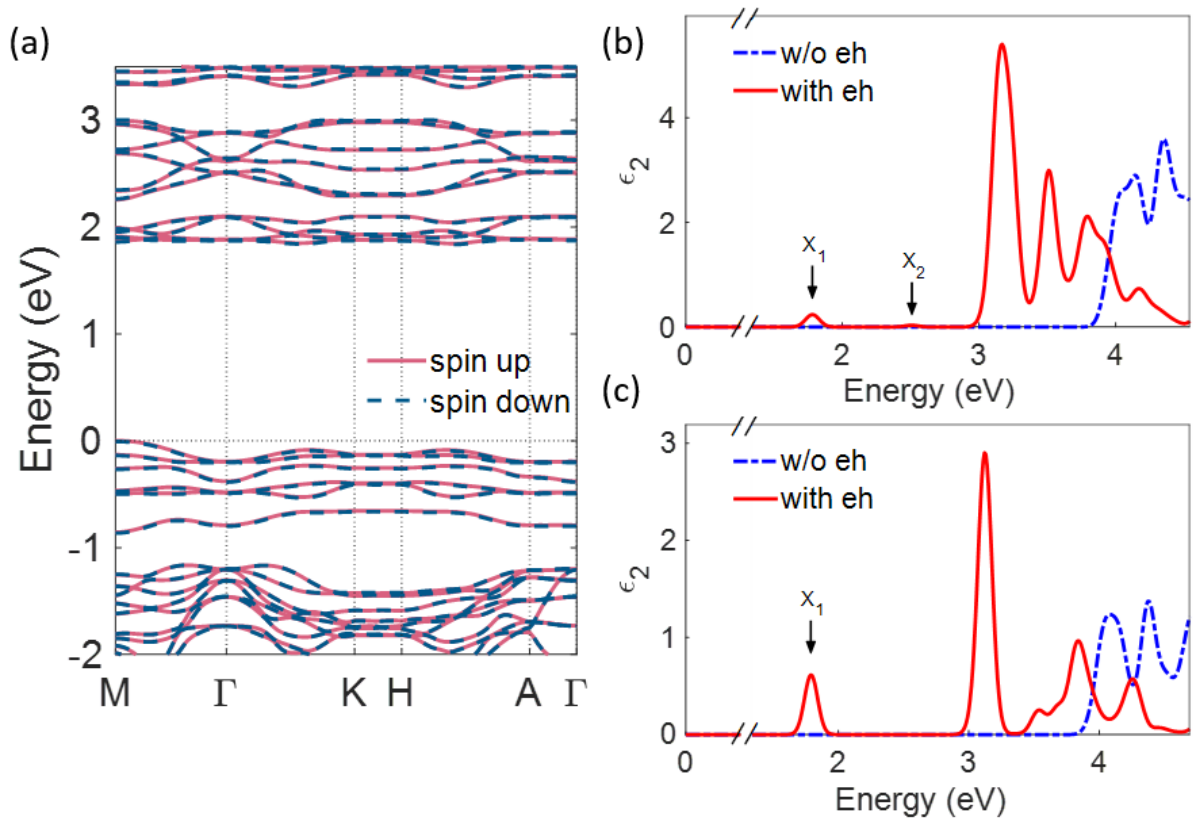

**Figure 7** Electronic and optical properties of bulk CrCl$_3$ in the LT rhombohedral stacking with an AFM interlayer coupling: (a) the DFT-calculated band structure. The energy of the valence band maximum is set to be zero. The optical absorption spectra without and with *e-h* interactions for incident light polarized in the (b) in-plane and (c) out-of-plane direction.

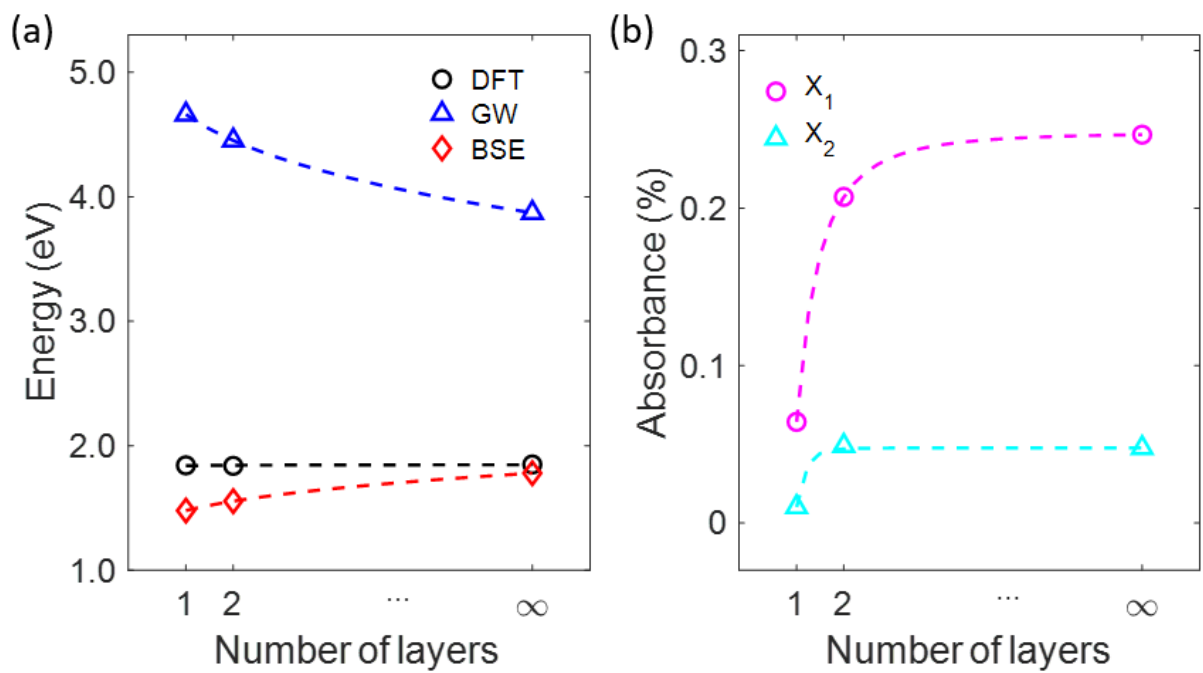

**Figure 8** (a) Evolution of DFT, QP band gaps and "optical gap" (energy of the first bright exciton $X_1$) of CrCl3 with the layer number. The dashed lines are power law fits to the results. (b) Evolution of the optical absorbance of the characteristic peaks $X_1$ and $X_2$ with the layer number.

**References:**

[1] K. F. Mak, C. Lee, J. Hone, J. Shan, and T. F. Heinz, Phys. Rev. Lett. **105**, 136805 (2010).

[2] K. He, N. Kumar, L. Zhao, Z. Wang, K. F. Mak, H. Zhao, and J. Shan, Phys. Rev. Lett. **113**, 026803 (2014).

[3] A. Chernikov, T. C. Berkelbach, H. M. Hill, A. Rigosi, Y. Li, O. B. Aslan, D. R. Reichman, M. S. Hybertsen, and T. F. Heinz, Phys. Rev. Lett. **113**, 076802 (2014).

[4] D. Y. Qiu, F. H. Da Jornada, and S. G. Louie, Phys. Rev. Lett. **111**, 216805 (2013).

[5] A. Ramasubramaniam, Phys. Rev. B **86**, 115409 (2012).

[6] W. Song and L. Yang, Phys. Rev. B **96**, 235441 (2017).

[7] T. Cheiwchanchamnangij and W. R. L. Lambrecht, Phys. Rev. B **85**, 205302 (2012).

[8] X. Wang, A. M. Jones, K. L. Seyler, V. Tran, Y. Jia, H. Zhao, H. Wang, L. Yang, X. Xu, and F. Xia, Nat. Nanotechnol. **10**, 517 (2015).

[9] G. Zhang, A. Chaves, S. Huang, F. Wang, Q. Xing, T. Low, and H. Yan, Sci. Adv. **4**, eaap9977 (2018).

[10] V. Tran, R. Soklaski, Y. Liang, and L. Yang, Phys. Rev. B **89**, 235319 (2014).

[11] V. Tran, R. Fei, and L. Yang, 2D Mater. **2**, 44014 (2015).

[12] N. Saigal, V. Sugunakar, and S. Ghosh, Appl. Phys. Lett. **108**, 132105 (2016).

[13] R. Schuster, J. Trinckauf, C. Habenicht, M. Knupfer, and B. Büchner, Phys. Rev. Lett. **115**, 026404 (2015).

[14] Y. Pan, S. Gao, L. Yang, and J. Lu, Phys. Rev. B **98**, 085135 (2018).

[15] M. L. Cohen and S. G. Louie, *Fundamentals of Condensed Matter Physics*. (Cambridge University Press, Cambridge, 2016).

[16] M. W. Lin, H. L. Zhuang, J. Yan, T. Z. Ward, A. A. Puretzky, C. M. Rouleau, Z. Gai, L. Liang, V. Meunier, B. G. Sumpter, P. Ganesh, P. R. C. Kent, D. B. Geohegan, D. G. Mandrus, and K. Xiao, J. Mater. Chem. C **4**, 315 (2016).

[17] C. Gong, L. Li, Z. Li, H. Ji, A. Stern, Y. Xia, T. Cao, W. Bao, C. Wang, Y. Wang, Z. Q. Qiu, R. J. Cava, S. G. Louie, J. Xia, and X. Zhang, Nature (London) **546**, 265 (2017).

[18] B. Huang, G. Clark, E. Navarro-Moratalla, D. R. Klein, R. Cheng, K. L. Seyler, Di. Zhong, E. Schmidgall, M. A. McGuire, D. H. Cobden, W. Yao, D. Xiao, P. Jarillo-Herrero, and X. Xu, Nature (London) **546**, 270 (2017).

[19] M. A. McGuire, G. Clark, S. Kc, W. M. Chance, G. E. Jellison, V. R. Cooper, X. Xu, and B. C. Sales, Phys. Rev. Mater. **1**, 014001 (2017).

[20] N. D. Mermin and H. Wagner, Phys. Rev. Lett. **17**, 1133 (1966).

[21] J. L. Lado and J. Fernández-Rossier, 2D Mater. **4**, 035002 (2017).

[22] B. Huang, G. Clark, D. R. Klein, D. MacNeill, E. Navarro-Moratalla, K. L. Seyler, N. Wilson, M. A. McGuire, D. H. Cobden, D. Xiao, W. Yao, P. Jarillo-Herrero, and X. Xu, Nat. Nanotechnol. **13**, 544 (2018).

[23] S. Jiang, J. Shan, and K. F. Mak, Nat. Mater. **17**, 406 (2018).

[24] D. Zhong, K. L. Seyler, X. Linpeng, R. Cheng, N. Sivadas, B. Huang, E. Schmidgall, T. Taniguchi, K. Watanabe, M. A. McGuire, W. Yao, D. Xiao, K. M. C. Fu, and X. Xu, Sci. Adv. **3**, e1603113 (2017).

[25] M. Wu, Z. Li, T. Cao, and S. G. Louie, Nat. Commun. **10**, 1 (2019).

[26] A. Molina-Sánchez, G. Catarina, D. Sangalli, and J. Fernández-Rossier, arXiv:1912.01888.

[27] M. A. McGuire, H. Dixit, V. R. Cooper, and B. C. Sales, Chem. Mater. **27**, 612 (2015).

[28] X. Cai, T. Song, N. P. Wilson, G. Clark, M. He, X. Zhang, T. Taniguchi, K. Watanabe, W. Yao, D. Xiao, M. A. McGuire, D. H. Cobden, and X. Xu, Nano Lett. **19**, 3993 (2019).

[29] H. H. Kim, B. Yang, S. Li, S. Jiang, C. Jin, Z. Tao, G. Nichols, F. Sfigakis, S. Zhong, C. Li, S. Tian, D. G. Cory, G. X. Miao, J. Shan, K. F. Mak, H. Lei, K. Sun, L. Zhao, and A. W. Tsen, Proc. Natl. Acad. Sci. **116**, 11131 (2019).

[30] X. Lu, R. Fei, and L. Yang, arXiv:2002.05208.

[31] Z. Sun, Y. Yi, T. Song, G. Clark, B. Huang, Y. Shan, S. Wu, D. Huang, C. Gao, Z. Chen, M. McGuire, T. Cao, D. Xiao, W. T. Liu, W. Yao, X. Xu, and S. Wu, Nature (London) **572**, 497 (2019).

[32] W. Song, R. Fei, and L. Yang, arXiv:2003.06385.

[33] J. P. Perdew, K. Burke, and M. Ernzerhof, Phys. Rev. Lett. **77**, 3865 (1996).

[34] S. Grimme, J. Comput. Chem. **27**, 1787 (2006).

[35] D. R. Hamann, Phys. Rev. B **88**, 085117 (2013).

[36] Supplementary information.

[37] J. Deslippe, G. Samsonidze, D. A. Strubbe, M. Jain, M. L. Cohen, and S. G. Louie, Comput. Phys. Commun. **183**, 1269 (2012).

[38] M. S. Hybertsen and S. G. Louie, Phys. Rev. B **34**, 5390 (1986).

[39] M. Rohlfing and S. G. Louie, Phys. Rev. B **62**, 4927 (2000).

[40] C. D. Spataru, S. Ismail-Beigi, L. X. Benedict, and S. G. Louie, Phys. Rev. Lett. **92**, 077402 (2004).

[41] L. Yang, C. D. Spataru, S. G. Louie, and M. Y. Chou, Phys. Rev. B **75**, 201304 (2007).

[42] H. Wang, V. Eyert, and U. Schwingenschlögl, J. Phys. Condens. Matter **23**, 116003 (2011).

[43] M. Cococcioni and S. De Gironcoli, Phys. Rev. B **71**, 035105 (2005).

[44] T. Miyake, P. Zhang, M. L. Cohen, and S. G. Louie, Phys. Rev. B **74**, 245213 (2006).

[45] B. C. Shih, Y. Xue, P. Zhang, M. L. Cohen, and S. G. Louie, Phys. Rev. Lett. **105**, 146401 (2010).

[46] L. Yang, J. Deslippe, C. H. Park, M. L. Cohen, and S. G. Louie, Phys. Rev. Lett. **103**, 186802 (2009).

[47] D. R. Klein, D. MacNeill, Q. Song, D. T. Larson, S. Fang, M. Xu, R. A. Ribeiro, P. C. Canfield, E. Kaxiras, R. Comin, and P. Jarillo-Herrero, Nat. Phys. **15**, 1255 (2019).

[48] I. Pollini and G. Spinolo, Phys. Status Solidi **41**, 691 (1970).

[49] P. Gu, Q. Tan, Y. Wan, Z. Li, Y. Peng, J. Lai, J. Ma, X. Yao, S. Yang, K. Yuan, D. Sun, B. Peng, J. Zhang, and Y. Ye, ACS Nano **14**, 1003 (2020).

[50] X. Zhao, C. M. Wei, L. Yang, and M. Y. Chou, Phys. Rev. Lett. **92**, 236805 (2004).